# A multi-center prospective evaluation of THEIA™ to detect diabetic retinopathy (DR) and diabetic macular edema (DME) in the New Zealand screening program

Ehsan Vaghefi[1,2]*, Song Yang[1], Li Xie[1], David Han[1], David Squirrell[1,3]

*Corresponding author

1-    Toku Eyes®, Auckland, New Zealand

2-    School of Optometry and Vision Science, The University of Auckland

3-    Department of Ophthalmology, The University of Auckland

Short Title: Clinical decision support DR-AI for New Zealand

* Corresponding Author

Dr Ehsan Vaghefi

88 Tapu Rd, Huapai

Auckland

New Zealand, 0810

e.vaghefi@tokueyes.com

P: +64 21 102 7705

**Word count:** 3623

**Conflict of interest**

Ehsan Vaghefi and David Squirrell are co-founders of Toku Eyes®, which is a start-up out of The University of Auckland, looking into commercialization of this artificial intelligence system (THEIA™) in New Zealand. No other conflict of interest for co-authors.

**Novelty Statement**

-       THEIA is proven to be the most accurate algorithm of its kind, through a double-blind prospective multi-centre trial

-       THIEA provides the highest level of disease diagnosis granularity, which is essential for early detection and timely intervention

-       THEIA provides an automated decision rule to ensure rapid, accurate classification of the large proportion of normal images from the few with abnormal features for prompt, accurate clinical grading, but not to replicate a screening program.

**Acknowledgments**

We wish to acknowledge the Diabetes eye screening program of CMDHB and Naylor Palmer optometrists for hosting the trial

**Authors contribution**

E.V. proposed the original research, developed the original methodology and developed the first draft of the manuscript. S.Y. , D.H & L.X. (equal contribution) supervised the technical aspect of the trial. D.S. supervised the clinical aspect of the trial and finalized the manuscript.

**Funding**

This work was funded by Ministry of Business, Innovation and Education of New Zealand (UOAX1805 - 3715780).

**Abstract**

Purpose: to assess the efficacy of THEIA™, an artificial intelligence for screening diabetic retinopathy in a multi-center prospective study. To validate the potential application of THEIA™ as clinical decision making assistant in a national screening program.

Methods: 902 patients were recruited from either an urban large eye hospital, or a semi-rural optometrist led screening provider, as they were attending their appointment as part of New Zealand Diabetic Screening programme. These clinics used a variety of retinal cameras and a range of operators. The de-identified images were then graded independently by three senior retinal specialists, and final results were aggregated using New Zealand grading scheme, which is then converted to referable\non-referable and Healthy\mild\more than mild\vision threatening categories.

Results: compared to "ground truth", THEIA™ achieved 100% sensitivity and [95.35%-97.44%] specificity, and negative predictive value of 100%. THEIA™ also did not miss any patients with "more than mild" or "vision threatening" disease. The level of agreement between the clinicians and the aggregated results was (k value: 0.9881, 0.9557, and 0.9175), and the level of agreement between THEIA™ and the aggregated labels was (k value: 0.9515).

Conclusion: Our multi-centre prospective trial showed that THEIA™ does not miss referable disease when screening for diabetic retinopathy and maculopathy. It also has a very high level of granularity in reporting the disease level. Since THEIA™ is being tested on a variety of cameras, operating in a range of clinics (rural\urban, ophthalmologist-led\optometrist-led), we believe that it will be a suitable addition to a public diabetic screening program.

# Introduction

Implementation of artificial intelligence (AI) in medicine and particularly in ophthalmology has a long history, but also accelerating rapidly in the past few years[1-4]. So far, the most promising application of AI in ophthalmology is as a screening tool for Diabetic Retinopathy (DR)[5-9].

It is now well accepted that a comprehensive DR screening program can reduce the burden of diabetes related vision loss[4, 10, 11]. However, delivering large community-based programs can be a major challenge even in developed countries, such as including New Zealand which has both a high prevalence of diabetes[12] and a significant proportion of the population not being screened regularly[13]. AI based algorithms, that can reliably detect DR in retinal images and provide instantaneous reporting with high diagnostic accuracy, could significantly improve the earlier detection of DR. In addition, by enabling specialist-level diagnostics to be provided to multiple peripheral sites simultaneously these algorithms also have the potential to massively increase access to, and lower the cost of, screening for DR[14, 15].

In recent years there have been significant advances in development of AI algorithms to assist with diabetic eye screening programs across the world[1]. While the accuracy of AI-based models for detecting DR have been demonstrated in many previous studies[5-9], most have failed to perform in the "real world" setting[16]. It has been shown that most research AIs for detection of retinopathy are not generalizable, as training datasets used are not representative of the wider society, obtained from relatively homogenous populations, limited in numbers or highly curated by clinicians, contain on image per eye, and very limited grade granularity (i.e. binary outcome for referable disease)[17]. Although there are several commercially available examples, very few AI products have thus far been integrated into a real-world clinical environment.

Here in New Zealand, Toku Eyes® has partnered up with local District Health Boards (DHB) to develop THEIA™, a AI DR Screening tool that is: trained and tested locally, is clinic\clinician\camera agnostic, gender\age\ethnicity unbiased, and provides retinopathy and maculopathy grading to the New Zealand Ministry of Health requirements[11]. In our recent retrospective study, we published results from the first iteration of THEIA™ which was trained and tested on a large dataset that represents 25% of New Zealanders living with diabetes. Since publication, the algorithm has been improved and optimised by a process of continuous retraining and testing. In this paper, we tested the latest iteration of THEIA™ in a multi-centre prospective trial, where the patients were recruited from two New Zealand National Diabetic Screening programs; a regional community Optometrist based provider and a Central Auckland DHB provider. Each program serves different communities using different cameras for their screening services. The aim of this study was to establish the efficacy of THEIA™, regardless of the type of fundus camera being used for, or location of the screening centre[5]. In this paper, we present the results of a bespoke AI that was developed to provide primary screening of diabetic retinopathy in order to augment the existing DR screening program in New Zealand; one that both accurately represents the real-world DR screening environment and is representative of the patients it is designed to serve.

# Methods

Study population: This was a prospective study, where participants were recruited from two separate clinics that are participating in the New Zealand Diabetic Screening program. One clinic was located in a large eye hospital setting in an urban area, while the other clinic was located in a semi-rural optometric practice. The central DHB service used a variety of Canon DGi (2 units), Canon CR2 (2 units) and Canon CR2+AF (2 units) cameras at its different sites, while the optometric led centre was using an iCare EIDON camera. The study protocol was approved by the Health and Disability Ethics committee at New Zealand Health and Disability Ethics Committee (20/STH/178). The trial is registered on the ANZCTR, Registration number ACTRN12620000488909 and has been issued with the Universal Trial Number (UTN) U1111-1249-7630.

We did not control for participants' ethnicity or gender, and kept the recruitment as 'first in, first served', aiming for a representative dataset of the New Zealand diabetic eye screening that we have published on before[14]. To ensure that there was a sufficient number of patients with diseased images, the study remained open until the desired number of patients with disease had been recruited. All patients attending for a publicly funded retinal screening (within the eye hospital or optometric setting) between January 2021 and April 2021, over the age of 17, were invited to participate. The only exclusion was vulnerable patients who were unable to give their consent.

The process of DR screening in New Zealand has been outlined previously, but in brief all participants with Type 2 diabetes mellitus (T2DM) were photographed twice in each eye, i.e. one macula-centered and one disk-centered image, and all patients with Type 1 diabetes mellitus (T1DM) were imaged four times in each eye, with an addition two images taken one below the disc and one above the disc. All patients are initially photographed through undilated pupils, pupil dilation being used if the image that was subsequently acquired was deemed by the photographer to be inadequate. At the conclusion of data collection, the images were de-identified and assigned a unique patient ID by an independent technician.

The de-identified images were then passed on to three independent specialists (senior graders in the three metro DHB screening programs). Each specialist graded the entire dataset independently, according to New Zealand Ministry of Health standards[11], and were masked to the grades issued by the others. Finally, a master ground truth list was created by aggregating the three independent reports. Where there was a discrepancy in the grades issued by the three independent graders a fourth independent, senior retinal specialist was used to adjudicate the outcome. This adjudicated data set formed the "ground truth" against which THEIA™ was assessed.

*Table 1: Aggregation of New Zealand diabetic screening standard grades to referable and non-referable disease gradings*

|  | Retinopathy | Maculopathy | Retinopathy & Maculopathy |
|---|---|---|---|
| Non-referable disease | R0, R1, R2 | M0, M1, M2 | R & M <3 |
| Referable disease | R3, R4, R5 | M3, M4, M5 | R or M >= 3 |

*Table 2: Aggregation of New Zealand diabetic screening standard grades to the international (Healthy, mild, more-than-mild, vision threatening) disease gradings*

|  | Retinopathy | Maculopathy | Retinopathy & Maculopathy |
|---|---|---|---|
| Healthy | R0 | M0 | R & M = 0 |
| Mild | R1, R2 | M1, M2 | R & M < 3 |
| mtmDR | R3 | M3 | R or M = 3 |
| Vision threatening | R4, R5 | M4, M5 | R or M > 3 , R & M = 3 |

Independent to the human grading pathway, the de-identified images were analyzed by the THEIA™ AI platform, by means of uploading images to its dedicated Amazon Web Services (AWS) portal. The THEIA™-generated grades were then compared with the ground truth. The efficacy of THEIA™ was assessed at the patient-level, based on the New Zealand grading system[18], as well as the basic binary referrable / non-referable [Table 1], and the more global (aggregated) grading scheme of Healthy, mild, more-than-mild (mtmDR) and vision threatening [Table 2]. Where a discrepancy existed between the results issued by THEIA™ and the adjudicated ground truth, the images were reassessed by the group who, being masked to the origin of the results, were asked to either agree with one of the outcomes presented.

# Statistical power calculation

Study success was pre-defined as both sensitivity and specificity of the AI system in the New Zealand population. The hypotheses of interest are H0: $p<p_0$ vs: HA: $p > p_0$ where p is the sensitivity or specificity of the AI system and $p_0 = 75\%$ for the sensitivity endpoint and $p_0 = 77.5\%$ for the specificity endpoint under the null hypotheses.

The alternative hypotheses were 85% for sensitivity and 82.5% for specificity, reflecting anticipated enrolment numbers and pre-specified regulatory requirements. One-sided testing was further prespecified for both sensitivity and specificity; a one-sided 2.5% Type I error was used resulting in a one-sided 97.5% rejection rule per hypothesis. To preserve Type I error, study success was defined as requiring both null hypotheses to be rejected at the end of the study.

Sample sizes for these hypotheses were calculated for at least 85% power and one-sided 2.5% Type 1 error requiring samples of at least 149 participants with referable DR or DME and 682 with non-referable retinopathy or maculopathy.

# Results

At the time of the study, and to address the large backlog that had resulted from the COVID lockdowns in New Zealand, the diabetic eye screening programs were prioritizing high risk patients with historically poor control or established retinopathy. We anticipated that this would result in recruiting a higher number of patients with disease than would usually be the case. Images were read sequentially but because there was both a time lag between the dates the results were issued and the date the patient was recruited and multiple sites were involved, more patients were recruited than the power calculation required (total 1050). Of these 246 individuals had disease that was deemed to represent referrable disease. Of these 246 patients, 2 had previously treated proliferative DR with extensive panretinal photocoagulation and in 1 there was an insufficient set of images to be accurately graded. These three participants were therefore excluded for the final analysis. The remaining patients had none or minimal disease and this dataset was randomly trimmed down to 597 patients to make a total of 902 patients whose images went forward to the studies grading team. The final calculations were therefore based on a total of 902 patients [Figure 1].

Using a simplified 2 class classification Referrable (R0-R2, M0-M2) v Non referrable (R3-R5, M4-M5), at **Patient-level** and using the worst R or M outcome in either eye, THEIA™ achieved a 100% sensitivity and 98.18% specificity, with the overall accuracy of 98%, [Table 3]. An extremely high level of agreement was observed between the three individual graders and the gold standard, (k values: 0.98, 0.96, and 0.92 respectively for the 3 graders) and THEIA™ (k value: 0.95). When diagnosing referable disease. The accuracy for identifying referable disease was [98% - 99%] for retinal specialists and 99% for THEIA™.

The **Eye level** results for the 2-class classification, broken down by the individual R and M grade results are shown in Supplementary Tables 2, 3 & 4. The sensitivity and specificity of detecting referrable retinopathy was 98.6% and 92.5% respectively, the sensitivity and specificity of detecting referrable maculopathy was 94.8% and 88.8% respectively.

The **Patient-level** results; using the worst R or M outcome in either eye, measuring THEIA™ performance against the gold standard with a granular 4 class classification; None (R0, M0), Mild (R1,2; M1,2), more than mild (R3, M3) and Vision threatening (R4,5; M4,5) are illustrated in the confusion matrix [Table 4]. These reveal that THEIA™ had a tendency to marginally over grade the disease, but it did not miss any patients with mtmDR disease or worse. THEIA™ issued a lower grade than gold standard in just 1 patient all of whom were issued with a mtmDR grade when the ground truth was considered to be Vision threatening. The level of agreement between the three individual graders and the gold standard using the 4 class classification ranged from k value: 0.96- 0.75. The corresponding level of agreement between THEIA™ and the gold standard was k 0.79. The accuracy for identifying referable disease was [86% - 98%] for retinal specialists and 89% for THEIA™.

The **Eye level** results for the 4-class classification, broken down by the individual R and M grade results are shown in supplementary tables 4 and 5. THEIA™ accurately grade the level of retinopathy in 1395/ 1713 (81.4%) of eyes and accurately graded the level of maculopathy in 1526/1702 (89.7%) of eyes.

THEIA™ also demonstrated similar level of proficiency in identifying referable disease, in both central DHB patients, and optometrist led practice [Tables 5 & 6].

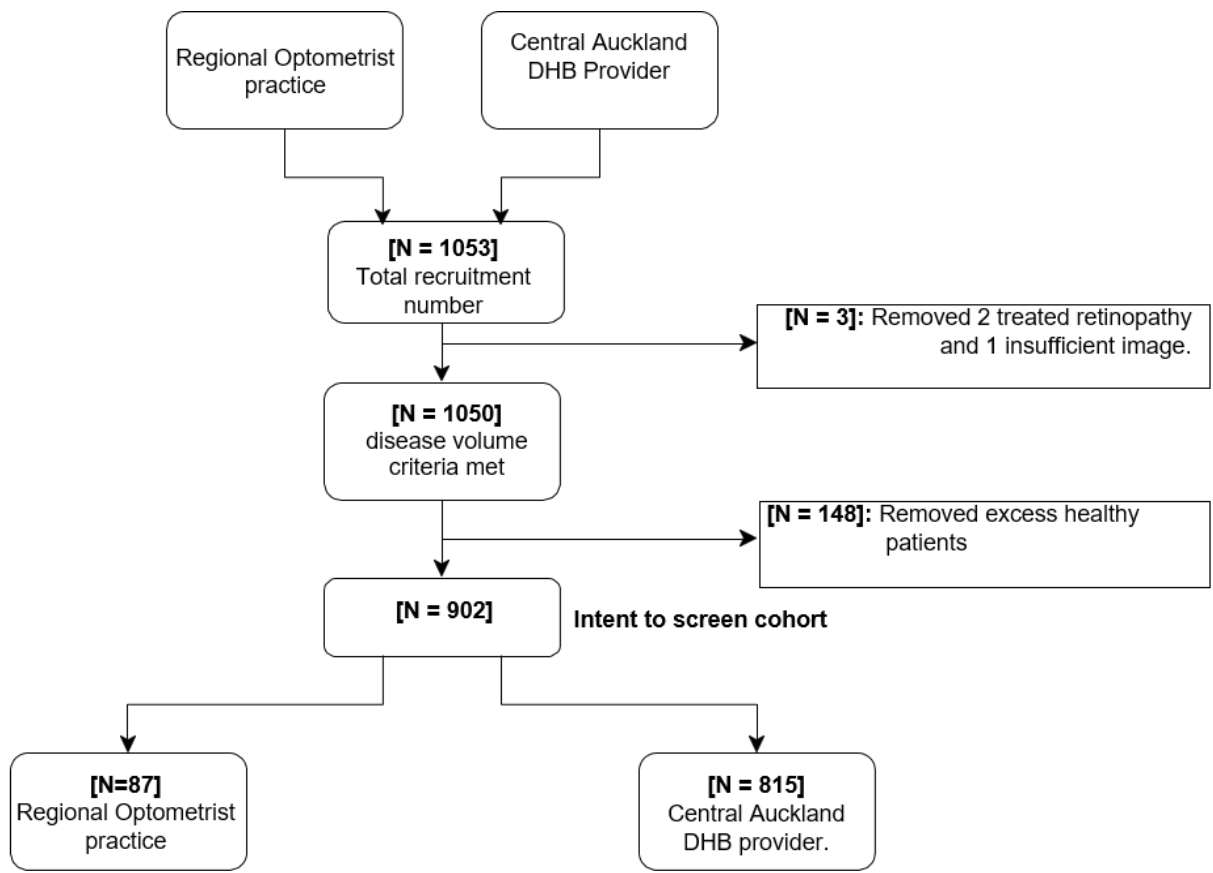

*Figure 1: The breakdown of participants analyzed for the THEIA V1 Clinical Study, from enrolment to analysis.*

Table 3: Patient-level results using 2 class assessment scale referral / non-referrable.

| OVERALL | ACCURACY | CONFUSION MATRIX | | SPECIFICITY | SENSITIVITY |
|---|---|---|---|---|---|
| CENTRAL AUCKLAND DHB | 98% | [599 [0 | 11] 205] | 98.19% | 100% |
| OPTOMETRIST LED PRACTICE | 99% | [49 [ 0 | 0] 38] | 100% | 100% |
| OVERALL | 98% | [648 [ 0 | 11] 243] | 98.33% | 100% |

Table 4: Patient-level THEIA™ confusion matrix using the (Healthy, Mild, mtmDR, Vision threatening) grading scheme.

| GOLD STANDARD\AI | HEALTHY | MILD | MTMDR | VISION THREATENING |
|---|---|---|---|---|
| HEALTHY | 343 | 123 | 0 | 1 |
| MILD | 22 | 160 | 2 | 8 |
| MTMDR | 0 | 0 | 17 | 30 |
| VISION THREATENING | 0 | 0 | 1 | 195 |

Table 5: Patient-level THEIA™ confusion matrix using the (Healthy, Mild, mtmDR, Vision threatening) grading scheme for Optometrist led practice .

| GOLD STANDARD\AI | HEALTHY | MILD | MTMDR | VISION THREATENING |
|---|---|---|---|---|
| HEALTHY | 21 | 10 | 0 | 0 |
| MILD | 5 | 13 | 0 | 0 |
| MTMDR | 0 | 0 | 4 | 3 |
| VISION THREATENING | 0 | 0 | 1 | 30 |

Table 6: Patient-level THEIA™ confusion matrix using the (Healthy, Mild, mtmDR, Vision threatening) grading scheme for Ophthalmologist led practice.

| GOLD STANDARD\AI | HEALTHY | MILD | MTMDR | VISION THREATENING |
|---|---|---|---|---|
| HEALTHY | 322 | 113 | 0 | 1 |
| MILD | 17 | 147 | 2 | 8 |
| MTMDR | 0 | 0 | 13 | 27 |
| VISION THREATENING | 0 | 0 | 0 | 165 |

# Audit of discordant grading

Using the 2 class classification system referrable v non-referrable, THEIA™ issued a different grade to the Gold standard adjudicated dataset in just 11 out of 902 patients. In all cases THEIA™ issued a grade that was higher than the gold standard. [Table 3, Figure 2]. Over grading of maculopathy resulted in 9 of the 11 cases where THEIA™ over graded compared to the gold standard. In 5 cases hard drusen were mistaken for exudate; Pachydrusen (2 cases) and thrombosed microaneurysms (2 cases) were responsible for the remainder. In 2 of the 11 cases where TEHIA™ disagreed with the gold standard, THEIA™ issued an R3 grade instead of an R2 grade. In both cases the retinopathy was at the R2/R3 interface The over-grading can be accounted for by the inbuilt "add up" function within THEIA™ which issued a patient level grade of R3 when the R grade classifier issued an R2 grade in both the disc and macular centred images. THEIA™ missed 4 cases (of 1713 eyes) where the eye level retinopathy was graded at mtmDR and 13 cases (of 1702 eyes) where the eye level maculopathy was graded as mtmDR.

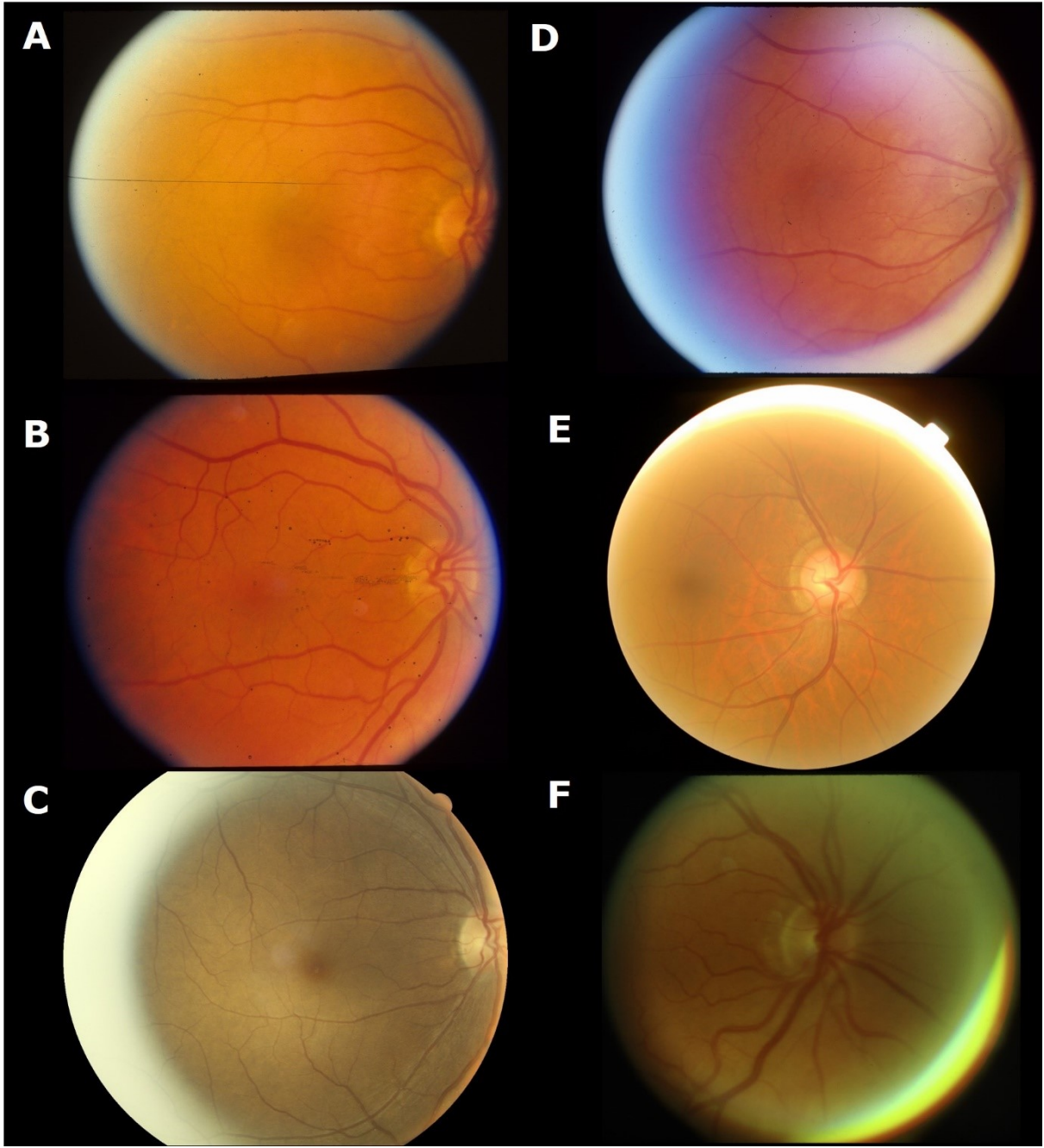

*Figure 2: Sample images of real-world image artifacts that was captured during this trial. THEIA has managed to use all of these images for grading. A) long dark horizontal strike through the image, B) thick reflection halo on the left hand side of the image, C) bubbles appear in the image, D) reflection halo surrounding the image, E) over exposure of the entire image, F) incorrect camera zoom and reflection halo*

## Other pathologies detected

In addition to screening for DR, our grading team was asked to comment on other sight threatening pathologies. Two patients in the cohort had a hemorrhagic branch retinal occlusion and one patient had a central retinal vein occlusion. Although THEIA™ was not able to identify these diseases specifically, all three were identified by THEIA™ as having "referrable" disease. No other sight threatening pathology was identified in this cohort.

# Discussion

While there has been a flurry of research designed to create artificial intelligence tools for screening diabetic retinopathy (DR) or diabetic macular edema (DME), very few algorithms have been tested prospectively in a real-world clinical environment[19-22]. In this study, we aimed to test the efficacy of our previously published algorithm (THEIA™) in a real-world prospective setting of two DR screening programs in New Zealand[5, 6]. We purposefully chose one urban DHB eye hospital screening center and one semi-rural optometrist led screening centre. We demonstrated that in a multi-centre prospective trial of more than 900 patients THEIA™ achieved 100% sensitivity, 98% specificity, with the overall accuracy of 98%. for identifying referable disease when compared to an adjudicated gold standard. At a more granular level of grading (Table 4), THEIA™ did not miss any cases of "mtmDR" or "Vision threatening" disease. The reasons for the few inconsistent grades between THEIA™ and the adjudicated gold standard were largely a result of drusen; both small hard drusen, and large pachydrusen, being mistaken for exudates. The gold standard adjudicated dataset was derived from grades issued by the senior grader in each of the three metro Auckland DHB screening programs. In keeping with their experience, the level of agreement between the individual graders and the adjudicated gold standard (k value: 0.92-0.98) when the data was aggregated into referrable v non referrable disease, was excellent. Although no cases of referrable disease were missed, all two of the human graders marginally under-graded compared to the adjudicated gold standard. This result was not statistically significant. A comparable level of agreement between THEIA™ and the adjudicated gold standard (k value: 0.95). In contrast to the human graders, THEIA™ marginally over-graded the images, a result which is in keeping with a tool which is designed with a high sensitivity and thus designed not to miss disease. Overall, these results demonstrate that THEIA™ is both reliable and is as consistent as experienced specialist graders in diagnosing and detecting referrable diabetic retinopathy and maculopathy in the New Zealand (or similar) screening program.

As expected, the accuracy of the level of agreement, for both the human graders and THEIA™ was less when they were assessed against the gold standard using a more granular grading system. In the case of THEIA™ the apparent drop off in performance, (k value: 0.78), can be explained by the fact that in grading DR we impose an ordinal scale onto a disease continuum. To reduce the likelihood of missing disease THEIA™ has been designed with an inbuilt tendency to over grade in situations where the disease sits at the boundary threshold of 2 disease states. Reassuringly THEIA™ accurately predicted the correct grade of retinopathy in 82% cases of retinopathy and 89% cases of maculopathy. In only 7 cases were retinopathy and maculopathy were considered together was vision threatening disease missed and in all cases THEIA™ labelled them as mtmDR.

Compared to other algorithms which have been assessed prospectively in a real world setting[7-9], THEIA™ performed very favorably. These results suggest that THEIA™ is not only the most accurate model of its kind, but is the only one capable of providing a very high granularity in the diagnosis of both retinopathy and maculopathy. Furthermore THEIA™, unlike other clinically tested AIs[23-26], provides these disease grades based on all images acquired per screening visit with the whole process from image acquisition through to grading being completely automated. While there has been significant interest in developing diabetic retinopathy grading AIs[27], few have been trained to specifically grade diabetic maculopathy as a separate entity[23-26], this despite diabetic maculopathy being the commonest reason for Ophthalmology referral[28]. Optical Coherence Tomography (OCT) is increasingly being incorporated into community screening programs and in addition to being highly accurate, THEIA™ is designed such that if it registers that sight threatening maculopathy may be present, the technician is prompted to perform an OCT scan before completing the imaging sequence.

THEIA™ has been designed primarily as a clinician assist primary triage tool. As such, it has been designed with an ultra-high sensitivity to ensure that sight threatening disease is not missed. In its previous configuration, THEIA™ achieved this at the expense of a modest specificity. With a modification to the algorithm, the current version of THEIA™ preserved its ultra-high sensitivity while achieving a specificity higher than 95%. In this trial, we used at least four different cameras in multiple clinical settings; one of which (The Optometrist practice) used an iCare Eidon camera (confocal scanning laser ophthalmoscopy technology), the other centres using a variety of Canon cameras (conventional flash photography technology). THEIA™'s performance was unaffected by the camera type used [Tables 5&6]. Furthermore, THEIA™ managed to cope with a number of artifacts on the real-world images, including a central bright halo that was generated by one camera, and a random assortment of dot artefacts that appeared in a consistent location in random images of a particular sequence from another camera [Figure 3].

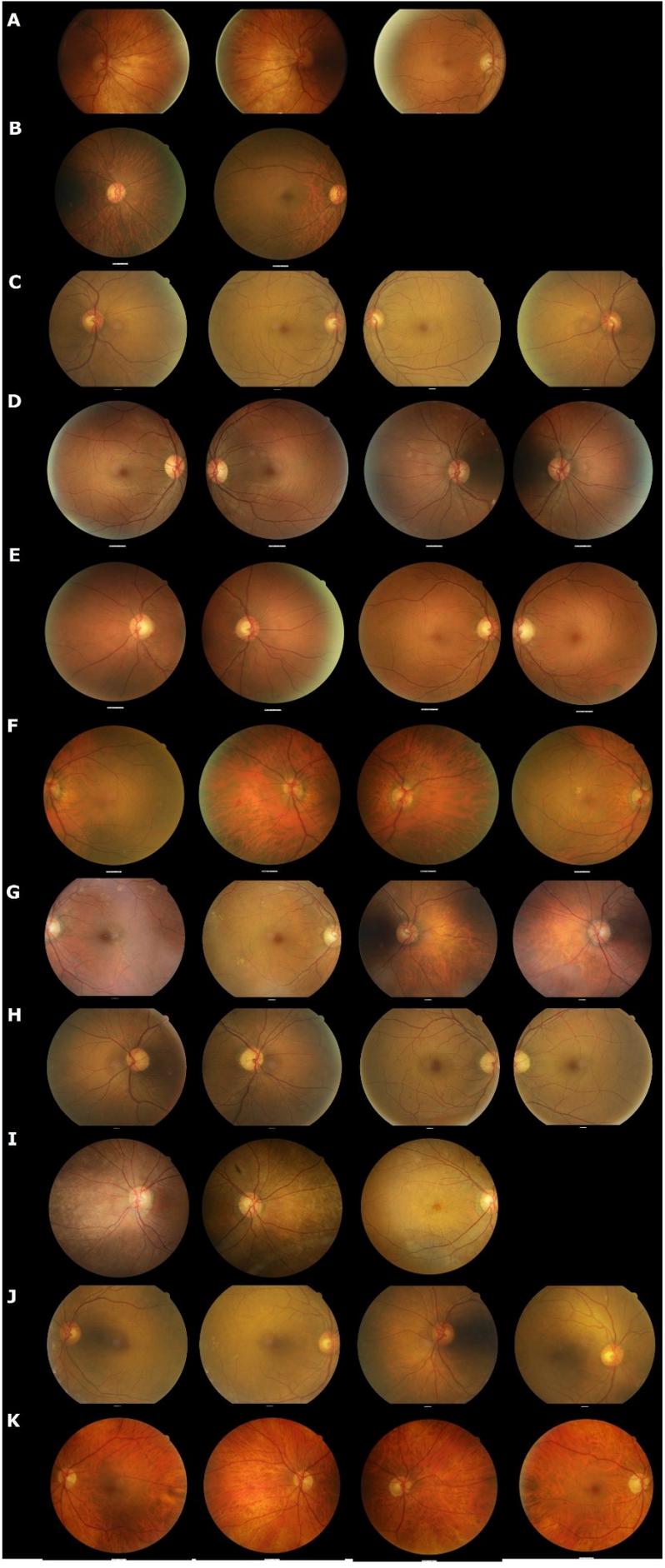

*Figure 3: 11 sets of image that were over-graded by THEIA™*

Whilst an accurate Algorithm is clearly important, there are a number of issues including equity and data privacy that have to be considered before AI can be safely incorporated into a diabetic screening program[17]. Since THEIA™ is designed to be compliant with these requirements, we believe that it will allow an established diabetic eye screening program (such as New Zealand's) to transition safely to a semi-automated system, where the clinicians capabilities (Graders, Optometrists and Ophthalmologists) are augmented, increasing overall screening capacity while reducing the overall cost. It has been shown that such a model is the most cost effective method of providing a national diabetic screening[29], with an estimated 20% cost saving compared to status quo.

The principal limitation of THEIA™ at this point of time is that it cannot reliably identify other eye diseases that might be present at the time of diabetic screening, such as glaucomatous optic neuropathy and age-related macular degeneration. Three patients in the current study had a significant retinal vein occlusion that was flagged up as significant retinopathy. It would also be reasonable to expect that hemorrhagic neovascular macular degeneration to be similarly identified. As a research group, we have for the past decade systematically collected all other pathologies that are detected during routine screening. Our analysis revealed (in press) that only severe hypertensive retinopathy, retinal vein occlusion and macular degeneration are sufficiently important to justify systematic detection during routine diabetic eye screening. Severe hypertensive retinopathy and many cases of advanced macular degeneration will already be picked up and flagged up by THEIA™ as mtmDR or Sight Threatening disease. Nevertheless, we are looking at incorporating an AI classifier that will be able to detect glaucoma suspects and intermediate and late AMD in addition to DR into future iterations of THEIA™.

In conclusion, this multi-centre prospective trial demonstrates that THEIA™ is one of the most accurate algorithms of its kind in detecting DR and DME, while providing a high level of granularity in grading. As such, and with appropriate oversight and audit, these results indicate that THEIA™ could be safely deployed within established diabetic screening programs to augment the expertise of the clinicians, increasing overall screening capacity while reducing costs per unit screen.

*Supplementary Table 1: Audit of discordant grades issued*

| | FINAL ADJUDICATED. | | | | THEIA PREDICTED | | | | DISCREPANCY | EXPLANATION |
|---|---|---|---|---|---|---|---|---|---|---|
| | Right eye | | Left eye | | Right eye | | Left eye | | | |
| PATIENT ID | R Grade | M Grade | R Grade | M Grade | R Grade | M Grade | R Grade | M Grade | | |
| TOKU-1030 | 2 | 1 | | | 3 | 1 | | | RE: R3 issued but R2 | R Just below threshold for R3. |
| TOKU-1072 | 1 | 2 | | | 0 | 4 | | | RE: M4 issued but M2 | Thrombosed MA |
| TOKU-1084 | 1 | 2 | 1 | 2 | 1 | 2 | 1 | 4 | LE: M4 issued but M2 | Hard drusen |
| TOKU-1120 | 0 | 0 | 0 | 0 | 0 | 0 | 0 | 4 | LE: M4 issued but M2 | Thrombosed MA |
| TOKU-1202 | 0 | 0 | 1 | 2 | 1 | 4 | 1 | 2 | RE: R1 issued but R0. LE:M4 issued but M0 | Thrombosed MA |
| TOKU-1258 | 0 | 0 | 1 | 0 | 1 | 4 | 3 | 0 | RE: M4 issued but M0. LE R3 issued but R1 | RE Pachy drusen mimicking exudate. LE: 2 blots left eye only |
| TOKU-1573 | 1 | 2 | 1 | 2 | 1 | 2 | 1 | 4 | LE: M4 issued but M0 | LE: Pachydrusen mimicking exudate |
| TOKU-1593 | 3 | 2 | 2 | 2 | 3 | 2 | 2 | 4 | LE: M4 issued but M2 | RE Hard drusen. |
| TOKU-1627 | 1 | 2 | | | 1 | 4 | | | RE: M4 issued but M2 | Linear Hard drusen |
| TOKU-1665 | 1 | 0 | 1 | 0 | 0 | 0 | 1 | 4 | LE: M4 issued but M0 | Linear Hard drusen |
| TOKU-1784 | 1 | 1 | 1 | 1 | 3 | 1 | 1 | 1 | RE: R3 issued but R1 | RE: Just below threshold R3. |

*Supplementary Table 1: Audit of discordant grades issued*

*Supplementary Table 2: Eye-level (Overall ref/non-ref) THEIA™ performance vs gold standard*

| RETINOPATHY | ACCURACY | CONFUSION MATRIX | | SPECIFICITY | SENSITIVITY |
|---|---|---|---|---|---|
| AUCKLAND CENTRAL DHB (DHB) | 98.3% | [1220 | 23] | 99.7% | 98.1% |
| | | [ 4 | 323] | 93.4% | 98.8% |
| OPTOMETRIST LED | 100.0% | [92 | 0] | 100.0% | 100.0% |
| | | [ 0 | 40] | 100.0% | 100.0% |
| OVERALL | 98.4% | [1312 | 23] | 99.7% | 98.3% |
| | | [ 4 | 363] | 94.0% | 98.9% |

*Supplementary Table 3: Eye-level (Retinopathy ref/non-ref) THEIA™ performance vs gold standard*

| RETINOPATHY | ACCURACY | CONFUSION MATRIX | | SPECIFICITY | SENSITIVITY |
|---|---|---|---|---|---|
| AUCKLAND CENTRAL DHB (DHB) | 98.3% | [1301 | 23] | 99.7% | 98.3% |
| | | [ 4 | 242] | 91.3% | 98.4% |
| OPTOMETRIST LED | 100.0% | [96 | 0] | 100.0% | 100.0% |
| | | [ 0 | 36] | 100.0% | 100.0% |
| OVERALL | 98.4% | [1397 | 23] | 99.7% | 98.4% |
| | | [ 4 | 278] | 92.4% | 98.6% |

*Supplementary Table 4: Eye-level (Maculopathy ref/non-ref) THEIA™ performance vs gold standard*

| MACULOPATHY | ACCURACY | CONFUSION MATRIX | | SPECIFICITY | SENSITIVITY |
|---|---|---|---|---|---|
| AUCKLAND CENTRAL DHB (DHB) | 97.5% | [1311 | 30] | 99.2% | 97.8% |
| | | [ 10 | 219] | 88.0% | 95.6% |
| OPTOMETRIST LED | 97.7% | [111 | 0] | 97.4% | 100.0% |
| | | [ 3 | 18] | 100.0% | 85.7% |
| OVERALL | 97.5% | [1422 | 30] | 99.1% | 97.9% |
| | | [ 13 | 237] | 88.8% | 94.8% |

*Supplementary Table 5: Eye-level retinopathy grade results*

| GROUND TRUTH \ AI | HEALTHY | MILD | MTMDR | VISION THREATENING |
|---|---|---|---|---|
| HEALTHY | 873 | 151 | 0 | 0 |
| MILD | 52 | 321 | 22 | 1 |
| MTMDR | 0 | 4 | 46 | 84 |
| VISION THREATENING | 0 | 0 | 1 | 147 |

*Supplementary Table 6: Eye-level maculopathy grade results*

| GROUND TRUTH \AI | HEALTHY | MILD | MTMDR | VISION THREATENING |
|---|---|---|---|---|
| HEALTHY | 1089 | 41 | 1 | 6 |
| MILD | 59 | 233 | 5 | 18 |
| MTMDR | 1 | 10 | 35 | 28 |
| VISION THREATENING | 0 | 2 | 5 | 169 |

*Supplementary Table 7: Eye-level combined retinopathy & maculopathy grade results*

| GOLD STANDARD \AI | HEALTHY | MILD | MTMDR | VISION THREATENING |
|---|---|---|---|---|
| HEALTHY | 868 | 217 | 2 | 6 |
| MILD | 28 | 207 | 16 | 60 |
| MTMDR | 0 | 3 | 27 | 62 |
| VISION THREATENING | 0 | 0 | 1 | 216 |